\documentclass[journal=jacsat,manuscript=article]{achemso}
\usepackage{dcolumn}
\usepackage{color}
\usepackage{ulem}
\usepackage{mathptmx, helvet, courier}
\usepackage{bm,amsmath,amssymb}
\usepackage{threeparttable}
\usepackage{multirow}
\usepackage{array}
\usepackage{etoolbox}
\usepackage{mathtools}
\usepackage{hyperref}
\usepackage{adjustbox}
\usepackage{chemformula} 
\usepackage[T1]{fontenc} 
\usepackage{multirow}

\usepackage{titlesec}
\titleformat{\paragraph}
{\normalfont\normalsize\bfseries}{\theparagraph}{1em}{}
\titlespacing*{\paragraph}{0pt}{3.25ex plus 1ex minus .2ex}{1.5ex plus .2ex}

\usepackage{comment}

\usepackage{tikz}
\usepackage{xcolor}

\title{Stabilizing $\alpha-$helicity of polypeptide in aqueous urea: Dipole orientation or hydrogen bonding?}

\author{Luis A. Baptista}
\affiliation{Max Planck Institute for Polymer Research, Ackermannweg 10, 55128 Mainz, Germany}
\author{Yani Zhao}
\affiliation{Max Planck Institute for Polymer Research, Ackermannweg 10, 55128 Mainz, Germany}
\author{Kurt Kremer}
\affiliation{Max Planck Institute for Polymer Research, Ackermannweg 10, 55128 Mainz, Germany}
\author{Debashish Mukherji}
\affiliation{Quantum Matter Institute, University of British Columbia, Vancouver BC V6T 1Z4, Canada}
\email{debashish.mukherji@ubc.ca}
\author{Robinson Cortes-Huerto}
\affiliation{Max Planck Institute for Polymer Research, Ackermannweg 10, 55128 Mainz, Germany}
\email{corteshu@mpip-mainz.mpg.de}

\date{\today}

\begin{document}

\setlength{\parindent}{0pt}

\begin{abstract}
Urea denatures proteins due to its strong tendency to dehydrate the first solvation shell via urea--residue preferential binding. However, even after extensive experimental and computational investigations, the influence of urea on the stability of secondary structures remains elusive. For example, contrary to the common understanding, experimental studies have indicated that specific polypeptides, such as poly--alanine or alanine--rich systems, may even show an improved tendency to form secondary structures in aqueous urea. We investigate this seemingly counter--intuitive behaviour 
using over 15$\mu$s long all-atom simulations. These results show how a delicate balance between the localized dipole orientations and hydrogen bonding dictates polypeptide solvation in aqueous urea. Our work establishes a structure--property relationship that highlights the importance of microscopic dipole--dipole orientations/interactions for the operational understanding of macroscopic protein solvation.
\end{abstract}

\maketitle
\section{Introduction}

Understanding (poly-)peptide solvation in solvents, especially in binary solvents, is at the onset of many developments in designing bio-compatible materials~\cite{Varanko_AnnRevBiomedEng22_343_2020,Dai_etal_BioMacroMol22_4956_2021,Haas_etal_FrontBioengBiotech10_2296_2022}. 
Interesting examples include, but are not limited to, responsiveness of polymers and hydrogels/microgels to external stimuli~\cite{Hydro2017,Mukherji20AR,ELHUSSEINY2022100186}, self-assembly of complex structures in solutions~\cite{Tritschler_Macromolecules50_3439_2017,Chiesa_BMCBiology18_35_2020}, and denaturation of proteins~\cite{Bathia_Udgaonkar_ChemRev122_8911_2022,Raghunathan_etal_BiophysRev12_65_2020,England_Haran_AnnuRevPhysChem62_257_2011}. In these cases, solubility, and thus structure, of a solute 
in water gets severely affected by the presence of osmolytes within the solvation shell, such as small alcohols, urea, and ions. Here, especially urea is a common osmolyte known to denature a native structure of a polypeptide sequence or a protein in general. Indeed, urea--peptide interaction is somewhat nontrivial given that many competing interactions govern the macroscopic solvation behavior \cite{Stumpe2007,Hua_PNAS105_16928_2008,Zangi_etal_JACS131_1535_2009}. In this context, most studies have discussed the 
importance of weak van der Waals (vdW) forces, the strength of which is less than a $k_{\rm B}T$ (or 2.48 kJ mol$^{-1}$) at room temperature $T = 300$ K, and relatively stronger hydrogen bonds (H$-$bond), the strength of which is between $4-8k_{\rm B}T$ (or 9.92$-$18.84~kJ mol$^{-1}$)~\cite{Mukherji20AR,hbond1,hbond2,hbond}. Here, $k_{\rm B}$ is the Boltzmann constant.\\
The generally accepted view of the microscopic origin of protein denaturation in aqueous urea states that the urea molecules preferentially interact with the protein backbone, impacting the hydrophobic core and thus the unfolding of a sequence~\cite{Hua_PNAS105_16928_2008,Zangi_etal_JACS131_1535_2009}. In this context, NMR experiments of alanine-rich polypeptides have suggested that the driving force for the urea--induced denaturation is due to a high tendency to form urea--residue H--bonds~\cite{Lim_etal_PNAS106_2595_2009} that disrupts the residue--residue (intramolecular) H--bond network responsible for stabilizing a helical conformation. Consistent with this common wisdom, circular dichroism (CD) spectroscopy results have shown that the $\alpha-$helix content, quantified in terms of mean residue ellipticity, gradually decreases with increasing urea molar concentration $c_{\rm u}$ below 4.0 M~\cite{Scholtz_etal_PNAS92_185_1995}. Ideally, if all residues denature in aqueous urea mixtures, one should expect a relatively rapid decrease of the $\alpha-$helix content with $c_{\rm u}$. However, the observed trends in CD (i.e., a weak initial decay in residue ellipticity) might indicate that-- while some residues tend to denature due to urea, specific residues may even fold under the influence of urea molecules. This observation is further supported by a recent set of experiments where it has been shown that a poly--alanine shows an increased degree of stable secondary structure in aqueous urea mixtures \cite{Hilser}.\\ 
%
The CD results discussed above \cite{Scholtz_etal_PNAS92_185_1995} were well reproduced by an \textit{effective} theoretical model. In particular, within the Zimm-Bragg description~\cite{Zimm_Bragg_JCP31_526_1959},
a polypeptide is approximated as a chain of residues, each in either a helix or a coil state. By computing the equilibrium constant of helix formation $s$, it is possible to estimate the theoretical mean residue ellipticity. The effect of urea can then be introduced using the linear extrapolation method (LEM), i.e., the shift in solvation (unfolding) free energy $\triangle G_{\rm s}$ is a linear function of $c_{\rm u}$ with a slope $m$.~\cite{Santoro_Bolen_Biochemistry27_8063_1988} This gives $s=s_{\circ}\exp(-m c_{\rm u}/RT)$, with $s_{\circ}$ being the value of $s$ at $c_{\rm u} = 0$ M and $R=8.314$ J mol$^{-1}$K$^{-1}$ is the gas constant. By fitting the experimental mean residue ellipticity, a value of 
$m=-0.12$ kJ mol$^{-1}$ M$^{-1}$ (or 0.048 $k_{\rm B}T$ M$^{-1}$) per residue was obtained~\cite{Scholtz_etal_PNAS92_185_1995}.\\
Here, we note in passing that a substantial contribution to the total free energy comes from H--bonds. Thus, the $m-$value estimated above is expected to be significantly larger, simply because of the
larger H--bond strength \cite{Mukherji20AR,hbond1,hbond2,hbond}. Indeed, simulations have reported over an order of magnitude larger $m-$value than the LEM predictions~\cite{Zhao:2020}. Therefore, the connection between these effective models, experiments, and the microscopic picture of the peptide solvation needs to be adequately clarified.\\  
Motivated by these observations, we investigate the structure--property relationship in polypeptide solvation to establish a correlation between microscopic interaction details and macroscopic conformations. In particular, we show how a delicate competition between dispersion, H--bonding, and local dipole--dipole interactions (DDI) controls the solvation behavior of a sequence. Here, it is essential to note that, to the best of our knowledge, simulations (to a large degree also experimental data) ignored the highly local DDI--based explanations, which are expected to play an essential and non--trivial role in 
the poly--peptide solvation. Indeed, a recent experimental work has commented on the importance of dipole orientations within FSS and its direct correlation with the solvation behavior \cite{Elsa_ACSCS8_1404_2022}.
For this purpose, we perform over 15$\mu$s molecular dynamic simulations by taking a model system of poly-alanine with sixty residues (Ala60) in aqueous urea mixtures (See Supplementary Figure S1 for the molecular structures).\\
The simulations are performed using the GROMACS molecular dynamics package \cite{gromacs}, where the CHARMM36m force field parameters \cite{CHARMM36m} are used to model an Ala60 and the TIP3P model~\cite{TIP1,TIP2} is used for the water molecules. 
Moreover, the CHARMM36m complemented force field parameters for the urea molecules are modified within the Kirkwood--Buff theory of solutions~\cite{KirkwoodBuff1951,Weerasinghe2003}. 
See Supplementary Section SII for more details. All simulations are performed under ambient conditions, i.e., at temperature $T = 298$ K and isotropic pressure $p = 1$ bar. Note also that
the estimates of energy in $k_{\rm B}T$ unit is calculated with respect to $T = 298$ K.

\section{Results and Discussions}

\begin{figure}[h!]
	\centering
	\includegraphics[width=1\textwidth]{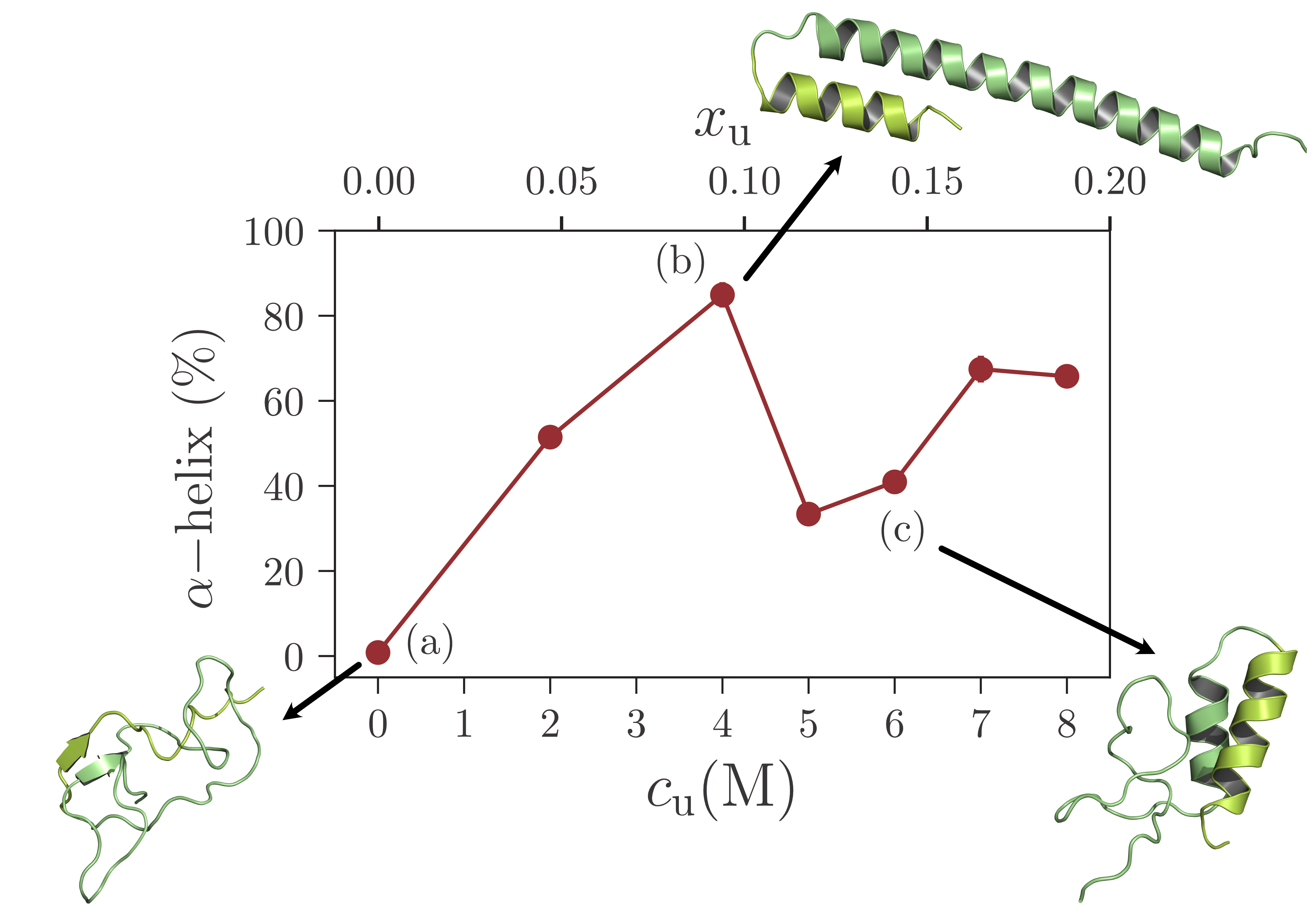}
	\caption{
		The percentage of $\alpha$-helix content as a function of urea mole concentration $c_{\rm u}$. An initial monotonic increase is observed with a maximum at $c_{\rm u}\simeq 4$ M 
		reaching 80$\%$ of $\alpha-$helical content in the peptide. A sharp subsequent decrease in helical content between 4 M$ <c_{\rm u}<6$ M signals at the denaturation of the protein. 
		Representative simulations snapshots at 0M, 4M and 6M illustrate the complex conformation behaviour with increasing urea concentration. Note for the clarity of
		presentation we have also included the urea mole fraction $x_{\rm u}$ in the upper panel of the abscissa. 
	} \label{Figure_01}
\end{figure}

\begin{figure}[h!]
	\centering
	\includegraphics[width=1.0\textwidth]{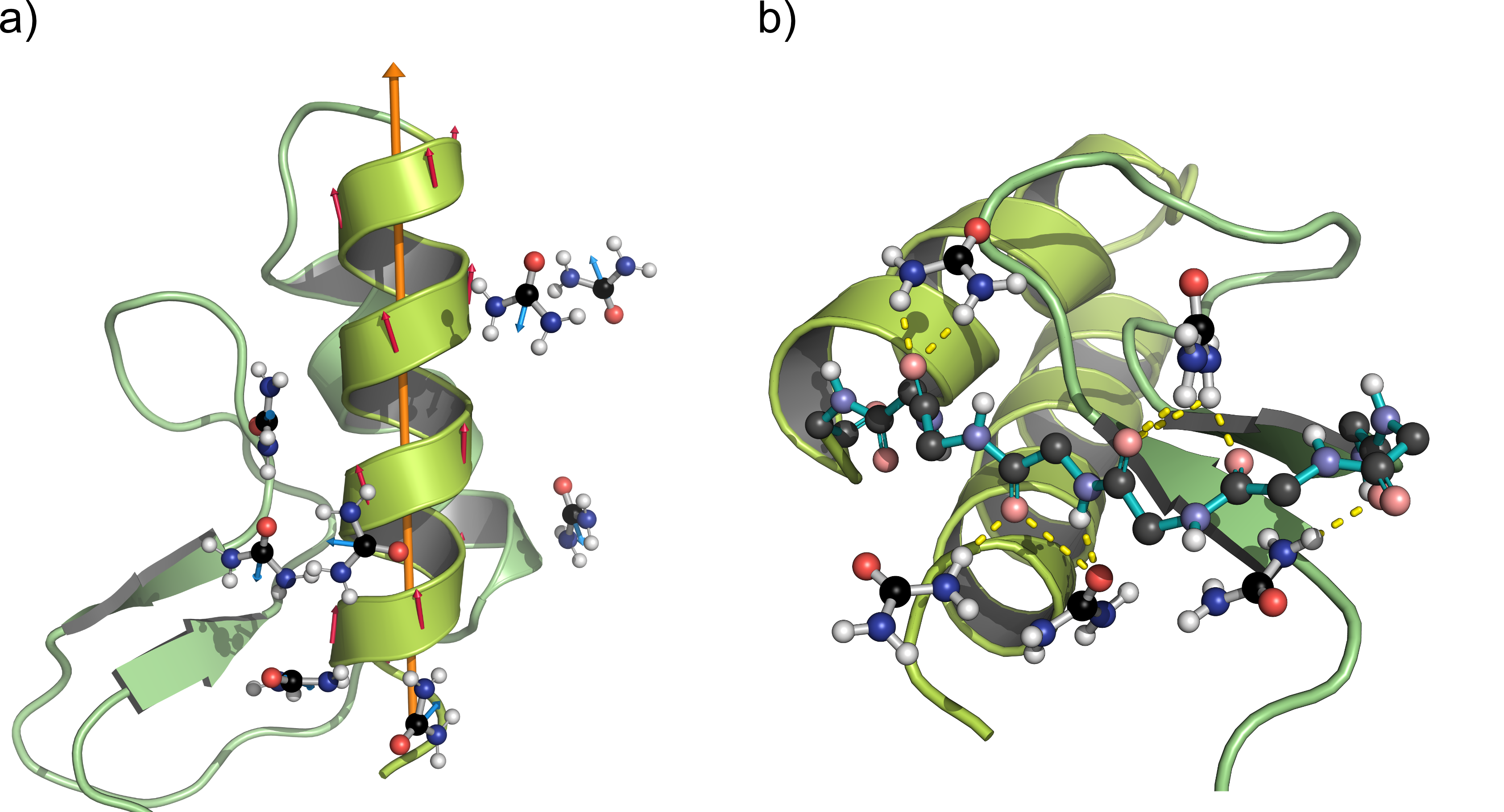}
	\caption{Simulation snapshots showing the detail of the orientation of urea molecules within the first solvation shell of Ala60 at 6 M. Panel (a) shows the urea molecules around 
	the $\alpha$-helix indicating urea electric dipoles (blue arrows), residue dipole (red arrows) and the total dipole of a helix (orange arrows). 
	Panel (b) shows the urea molecules forming hydrogen bonds (yellow dashed lines) with the peptide backbone of the loop region (blue bonds). 
	For clarity, the methyl group and hydrogen atoms are not visible in the representation. 
	}
	\label{Figure_02}
\end{figure}

In Fig. \ref{Figure_01}, we start our discussion by commenting on the $\alpha-$helix content as a function of $c_{\rm u}$. Note that for the clarity of presentation, we also report the urea mole fraction $x_{\rm u}$. It can be appreciated that the data show a non-monotonic functional dependence with $c_{\rm u}$. More specifically, different solvation regions are observed: (a) For $c_{\rm u} = 0.0$ (i.e., in pure water), we observe a rather globular conformation that expands upon increasing $c_{\rm u}$. (b) Starting at $c_{\rm u}\simeq 2$~M, the $\alpha-$helix content grows up to 60\%, which at $c_{\rm u} \simeq 4 $~M propagates almost to the whole Ala60 (see the simulation snapshots in Fig.~\ref{Figure_01}). This initial stabilization is somewhat in agreement with the experimental results on poly--alanine in aqueous urea \cite{Hilser} and may provide a possible explanation for the initial weak reduction in the degree of secondary structure \cite{Scholtz_etal_PNAS92_185_1995}. (c) Within the range 4 M$ < c_{\rm u} <$6 M, we observe an apparent destabilization of the $\alpha-$helix content, with Ala60 exhibiting relatively large coil regions. (d) Finally, there is a second $\alpha-$helix stabilization at around $c_{\rm u}=7$~M. Here, we note in passing that-- while the $\alpha-$helix content shows a non-monotonic variation, $\triangle G_{\rm s}$ decreases monotonically with $c_{\rm u}$ for polyalanine~\cite{Zhao:2020}. This difference suggests that $\triangle G_{\rm s}$ and the conformational behavior are somewhat decoupled. 
A structure--thermodynamic decoupling was also observed when a polymer \cite{mukherji13mac} or an intrinsically disordered polypeptide \cite{Zhao:2020} collapses in a mixture of two miscible solvents. However, the microscopic origin of polymer collapse (i.e., a spherical globule) is qualitatively different from the trends (i.e., stability of a secondary structure) observed in Fig.~\ref{Figure_01}.\\ 
What causes such a counter-intuitive structure--thermodynamic relationship? To address this issue, we will now start to investigate the alanine-(co-)solvent coordination within the first solvation shell (FSS). To this aim, we define the cylindrical correlation functions by employing the symmetry axis indicated in the helix. See the orange arrow in Fig.~\ref{Figure_02}(a). As expected, the data suggest an excess of urea molecules around the helix for all $c_{\rm u}$ (See Supplementary Figure S6), indicating the well-accepted concept of preferential binding of urea with polyalanine. Note also that the generally accepted microscopic origin of this urea preferential binding to a poly-peptide was shown to be dictated by the dispersion forces~\cite{Hua_PNAS105_16928_2008,Zangi_etal_JACS131_1535_2009}. However, if the dispersion forces were the only driving force in Ala60 solvation, $\triangle G_{\rm s}$ is expected to be much weaker than $6-8~k_{\rm B}T$ \cite{Zhao:2020}. Hence, other stronger interactions are also expected to be relevant.\\
A closer inspection of the urea molecules within FSS reveals an interesting molecular arrangement. See Figure~\ref{Figure_02}. It can be appreciated that the urea dipole moments (shown by the blue arrows in Figure \ref{Figure_02}(a)) tend to align with the dipole moment of the residue (shown by the red arrows in Figure \ref{Figure_02}(a)). Close to the coils, however, the urea molecules form H-bonds with Ala60, represented by yellow dashed lines in Figure \ref{Figure_02}(b).\\
To quantify the degree of alignment between Ala60 and urea dipole moments, we filter the cylindrical distribution function into dipole channels: parallel, antiparallel and "middle" relative orientations with respect to the $\alpha-$helix (See the Supplementary Figure S7). These results indicate an important alignment between the $\alpha-$helix and urea dipole moments in the FSS, with a clear preference for antiparallel alignment, for all $c_{\rm u}$. Within this picture, it is expected that the alignment of the dipoles corresponds to the energetically stable configuration that decreases the binding free energy of the system. Indeed, it has been revealed that molecular dipole-dipole interactions play an essential role in determining certain protein properties~\cite{Hol_etal_Nature271_443_1978,Jing_Quiocho_ProSci2_1643_1993,MilnerWhite_ProSci6_2477_1997}. However, the precise role of these interactions between the residues and the cosolvents remains elusive~\cite{Sippel_Quiocho_ProSci24_1040_2015}, with recent experimental efforts aiming at elucidating the electric dipole network in the FSS~\cite{Elsa_ACSCS8_1404_2022}.\\
A simple expression to estimate the energetic contribution due to the dipole interactions is the so-called Keesom energy $E_{\rm K}=-2\mu_{\rm r}^2\mu_{\rm u}^2/3\mathcal{D}^2k_{\rm B}Tr^6$~\cite{keesom1921van} with $\mathcal{D} = 4\pi\times$8.8541878$\times$10$^{-12}$~C~N$^{-1}$m$^{-2}$ the dielectric constant. Taking the average values of the dipole moments for a residue $\mu_{\rm r} = 3.95$~D and a urea molecule $\mu_{\rm u}=4.38$~D, we find $E_{\rm K} \simeq -6.55$ kJ mol$^{-1}$ (or 2.64 $k_{\rm B}T$)~\cite{keesom1921van} for $r = 0.6$ nm (i.e., range of the first peak in the Supplementary Figure S6) at $T=298$ K. This suggests that the dipole--dipole interactions (DDI) are indeed relevant. We also note in passing that $E_{\rm K}$ considers all possible orientations of the dipoles involved within $r=0.6$ nm. At the same time, the most dominant contributions usually come from the dipole orientations between the residues within an Ala60 (referred to as {\bf Scenario I}) and between a residue and the urea molecules (referred to as {\bf Scenario II}). In the following, we differentiate different contributions to the total resultant dipole--dipole interactions.\\
\begin{figure}[h!]
	\centering
	\includegraphics[width=1\textwidth]{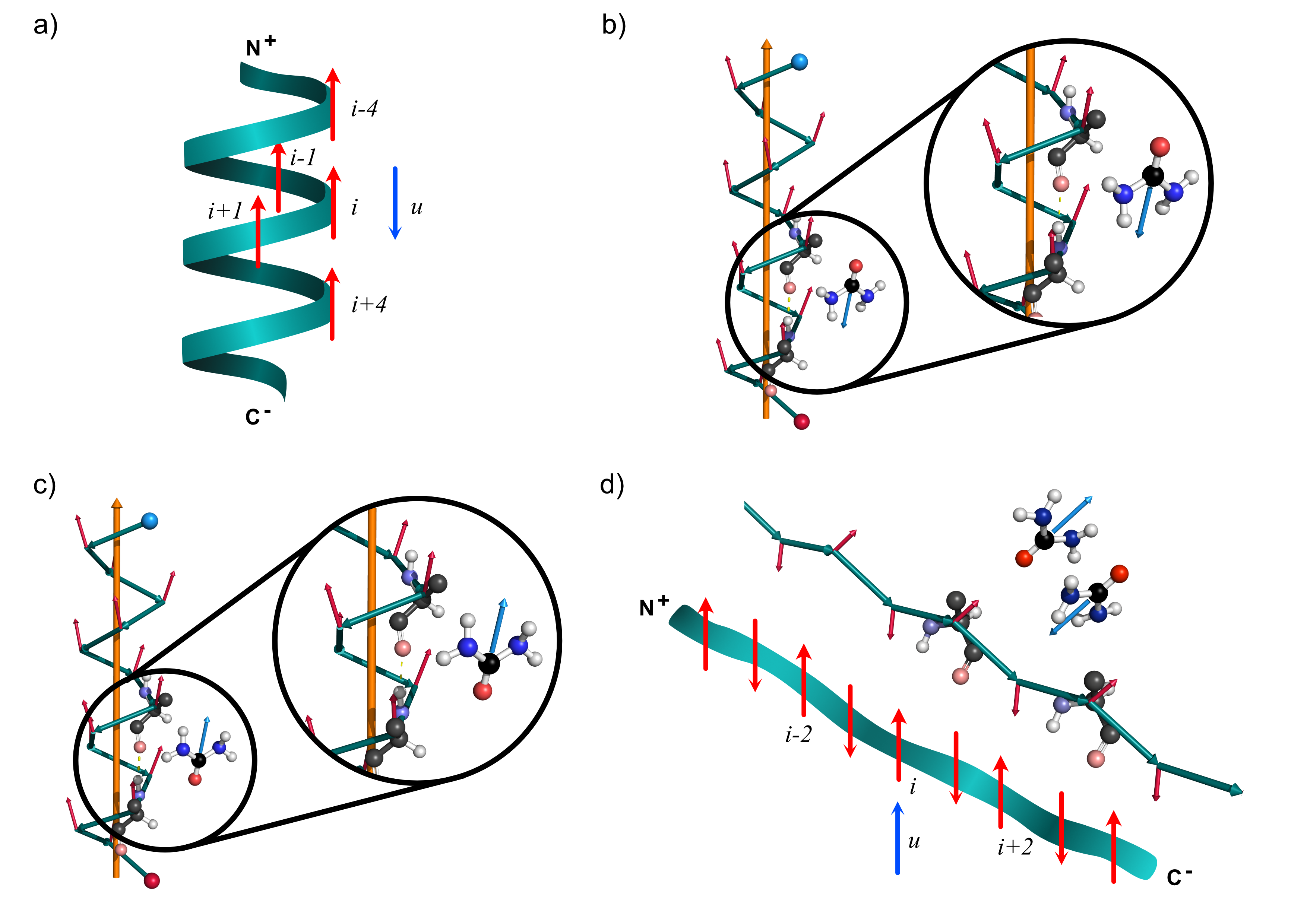}
	\caption{Schematic and ideal model representations of the possible dipole moment configurations in the system. 
	Panel (a) shows the $\alpha-$helix conformation with their consecutive residues having the dipoles aligned (red arrows) and with neighbouring urea molecules orienting their dipoles (blue arrows) side--by--side.
	Note that we show the most dominant relative orientation, i.e., the anti--parallel arrangement of an urea and the residues. Panel (b) represents the snapshot of an equivalent model system with an ideal $\alpha-$helix and an urea dipole moment located at $r = 0.6$ nm. Here, an anti-parallel side--by--side dipole arrangement between urea (blue arrow) and residue (red arrow) is shown. Panel (c) is same as panel (b), however, when an urea molecule is stabilised in between the $i$ and $i+4$ residues showing a parallel alignment. Panel (d) shows a fully unfolded structure is more stable from the point of view of dipole-dipole interactions, i.e., a locally dipole compensated arrangement. 
	The side chain Hydrogen atoms are not displayed for clarity. 
	}
	\label{Figure_03}
\end{figure}

\begin{figure}[ptb]
	\centering
	\includegraphics[width=0.5\textwidth]{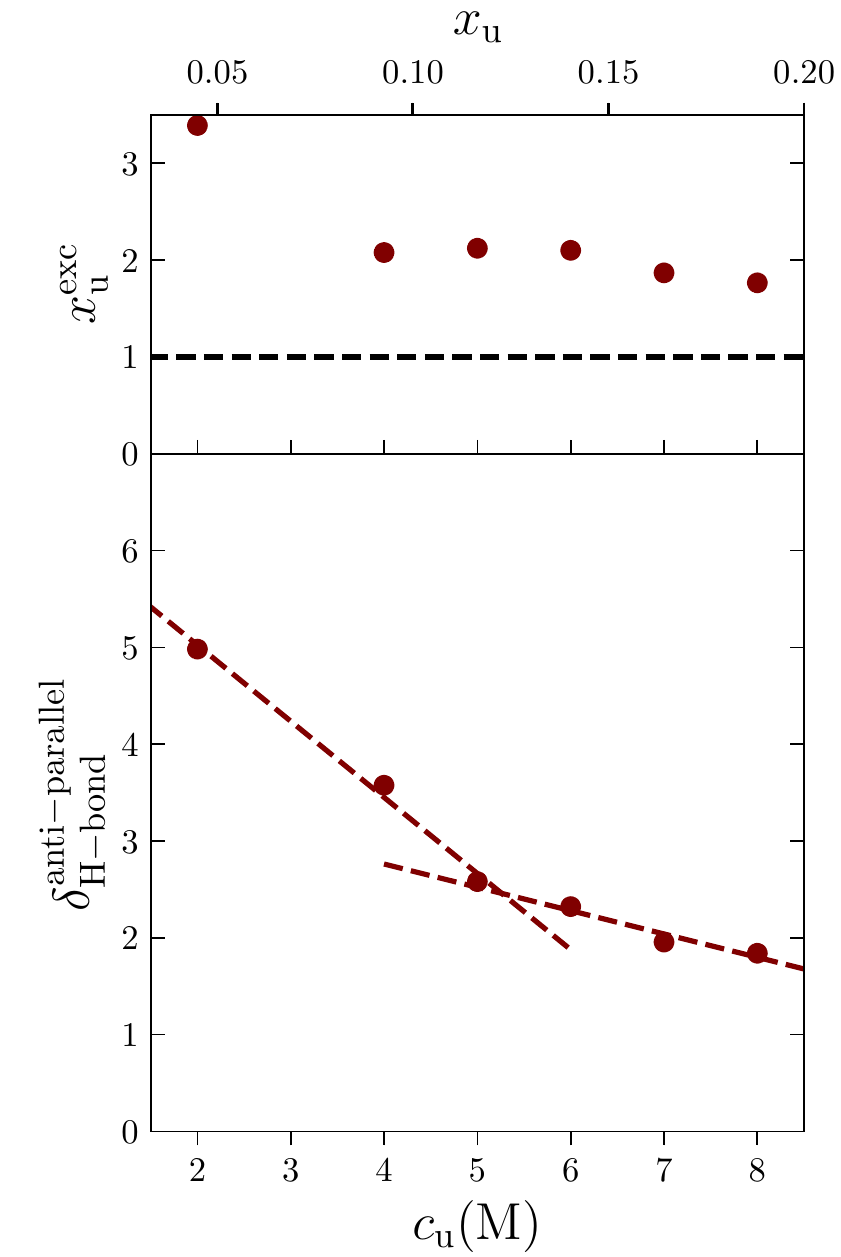}
	\caption{(Upper panel) Excess number of urea--peptide hydrogen bonds $x_{\rm u}^{\rm exc}$ as a function of urea mole concentration $c_{\rm u}$. The black line is a guide to the eye.
	The observed trend $x_{\rm u}^{\rm exc} > 1.0$ is in agreement with the mechanism of $\alpha-$helix unfolding reported experimentally~\cite{Lim_etal_PNAS106_2595_2009} and schematically depicted in Figure~\ref{Figure_03}. (Lower panel) Ratio between anti-parallel urea molecules in the first solvation shell and the number of urea-peptide H-bonds as a function of urea concentration. 
	Two linear regimes with a crossover between $c_{\rm u}=$4 and 5~M are apparent, in registry with Figure~\ref{Figure_01}. 
	DDI dominates the solvation behavior at low concentrations and H--bond becomes increasingly important at high concentration. 
	The lines are drawn to guide the eye.} \label{Figure_04}
\end{figure}

\textbf{Scenario I}: We first consider an $\alpha-$helix, shown by the model representation in Figures~\ref{Figure_03}(a-b). For this purpose, we build a perfect helix with 3.6 residues per turn and use the charge values given by the CHARMM36m force field parameters for the $\mu$ estimates.~\cite{CHARMM36m} We evaluate the dipole--dipole interaction energy on a residue $i$ using $E_{i} = \sum_{j\in \langle i,j \rangle} \varepsilon_{i,j}$~\cite{van2006principles}, where 
\begin{equation}
\label{eq:dmodel}
    \varepsilon_{i,j} = \frac{\bm{\mu}_i \cdot \bm{\mu}_{j}}{\mathcal{D}~|\mathbf{r}|^3} 
              - \frac{ 3 ( \bm{\mu}_i \cdot \mathbf{r} ) (\bm{\mu}_{j} \cdot \mathbf{r} ) }{\mathcal{D}~|\mathbf{r}|^5}\,
\end{equation}
and $\langle i,j\rangle$ represents the nearest neighbours of $i$, i.e., the residues $i\pm 4$, $i\pm 3$, $i\pm 2$ and $i\pm 1$. 
$\bm{\mu}_i$ is the electric dipole of the $i$-th residue and the vector $\mathbf{r} = \mathbf{r}_i - \mathbf{r}_j$ being the separation between the dipoles $i$ and $j$. Eq.~\ref{eq:dmodel} gives $E_i \simeq -5.29$ kJ~mol$^{-1}$ (or -2.13 $k_{\rm B}T$), indicating that, in addition to intra--molecular H--bonds, the $\alpha-$helix is likely to be stable via intra--molecular DDI.\\
\textbf{Scenario II}: 
One possible molecular arrangement is when a urea molecule stacks side--by--side in antiparallel dipolar orientation within FSS of the residue $i$. Using the same model above, but now including the urea molecule, we obtain $E_{\rm u} = \sum_{j\in \langle \rm{u}, j\rangle} \varepsilon_{\rm{u},j} \simeq -6.99$ kJ mol$^{-1}$ (or 2.81 $k_{\rm B}T$) for $r = 0.6$ nm separation between the geometric centers of the residue and the urea molecule. See Figure~\ref{Figure_03}(b) and the Supplementary Figure S8.\\ 
The second possible configuration is when a urea molecule sits between the residues $i$ and $i+4$, for example, with dipole moments aligned in head--to--tail orientation, 
with the head being the NH$_2$ groups of urea and the tail as carboxyl group of an alanine residue. See Figure~\ref{Figure_03}(c). In this case, we obtain $E_{\rm{u}} \simeq +1.89$ kJ mol$^{-1}$ (or +0.76 $k_{\rm B}T$). Since the latter estimate is smaller than $k_{\rm B} T$, the thermal fluctuations are expected to induce flip--flop of a urea dipole moment at this particular position and thus introduce local configurational fluctuations.\\ 
Even when the energy values discussed above suggest that the dipole interaction can stabilize an $\alpha-$helix, it is still important to emphasize that DDI alone are insufficient to stabilize the secondary structure.
For example, considering a test case of a fully expanded chain (shown in Figure~\ref{Figure_03}(d)), we find $E_i \simeq -15.95$ kJ mol$^{-1}$ (or -6.34 $k_{\rm B}T$) and $E_{\rm u} \simeq -20.65$ kJ mol$^{-1}$ (or -8.20 $k_{\rm B}T$) for an incoming urea molecule with dipole moment aligning head--to--tail with the $i$--th residue. See Supplementary Figure S9. These results indicate that, from the DDI point of view, the peptide unfolded state is more favorable than a stable $\alpha-$helix conformation by $\Delta E_{i} \simeq -10.66$ kJ mol$^{-1}$ (or 4.24 $k_{\rm B}T$). Hence, a pure dipole--based energetic argument can not fully explain the conformational behavior of an Ala60 in aqueous urea mixtures. In addition to DDI, we underline that $\alpha-$helix structures are further stabilized by H--bond interactions between $i$ and $i\pm 4$ residues. We will return to this point in the last part of the manuscript.\\
Given the discussion presented above, we can now understand the non-monotonic variation in $\alpha-$helicity with $c_{\rm u}$.
Here, we propose the following microscopic mechanics of the $\alpha-$helix stabilization:
\begin{itemize}

\item For $c_{\rm u} \le 4.0$ M (or below 10\%), i.e., when only a small amount of the urea molecules are added in the aqueous solution, they preferentially arrange in the side--wise antiparallel configurations next to the Ala60 residues because of the favorable minimum energy configuration ({\bf Scenario I}) and thus dehydrate FSS of Ala60. 
This scenario stabilizes an $\alpha-$helix structure by the urea--residue DDI, in addition to H--bond interactions between $i$ and $i\pm 4$ residues.

\item Upon further increase of the urea concentration, i.e., for $c_{\rm u} > 4$ M, in addition to the side--wise arrangements discussed above, the additionally urea molecules can also sit within the interstitial regions between $i$ and $i\pm 4$ residues ({\bf Scenario II}). In this region, urea dipoles can flip--flop and thus disturbs the intra-molecular H--bond network, promoting the formation of inter-molecular H--bonds between the NH$_2$ groups of urea and the carboxyl groups of Ala60. Moreover, this urea--residue arrangement is also favorable from the viewpoint of DDI. These combined effects then induce the $\alpha-$helix unfolding into a dipole--compensated structure within the range 4 M $\le c_{\rm u} \le$ 6 M, see Figure~\ref{Figure_01}. This interpretation is consistent with experiments indicating that the urea molecules affect the intra-molecular H--bond network~\cite{Lim_etal_PNAS106_2595_2009}, destabilizing the Ala60 secondary structure.

\item The apparent second stability region for $c_{\rm u} > 6.0$ M is somewhat surprising. We do not have a concrete argument for such behavior. However, based on the trajectory analysis, 
we speculate that the urea molecules within FSS join hands, making an H--bonded chain-like configuration that decreases the number of urea--residue H--bonds. Free of urea's strong influence, the C$_\alpha$ H--bond network reconnects, stabilizing the $\alpha-$helix.

\end{itemize}
We return to the H--bond analysis to further consolidate our arguments based on the competition between DDI and H--bonds. For this purpose, we have computed the excess number of hydrogen bonds $x_{\rm u}^{\rm{exc}} = x_{\rm u}^{\rm H-bond}/x_{\rm u}$ between urea and the peptide, where $x_{\rm u}^{\rm H-bond}$ is defined as,
\begin{equation}
x_{\rm u}^{\rm H-bond} = \frac{N_{\rm u-r}^{\rm H-bond}}{N_{\rm u-r}^{\rm H-bond}+N_{\rm w-r}^{\rm H-bond}}\, ,
\end{equation}
with $N_{\rm u-r}^{\rm H-bond}$ and $N_{\rm w-r}^{\rm H-bond}$ being the average number of H--bonds between residue and urea and residue and water, respectively. 
$x_{\rm u}^{\rm{exc}}>1$ indicates an excess of H-bonds, as seen in the upper panel of Figure~\ref{Figure_04}. This result demonstrates that urea strongly tends to 
make H-bonds with Ala60, in good agreement with the NMR data~\cite{Lim_etal_PNAS106_2595_2009}. This result also agrees with the statement that increasing urea--residue H--bonds disturbs the C$_\alpha$ H--bond network that drives the overall destabilisation of the $\alpha-$helix structure.\\
Lastly, to unravel the competition between dipole alignment and H-bond formation with urea concentration, we focus on the average number of urea ($N_{\rm u-r}^{\rm antiparallel}$) and water ($N_{\rm w-r}^{\rm antiparallel}$) molecules in the FSS with antiparallel dipole orientation ($-1\le \langle\cos\theta\rangle<-0.98$, see the Supplementary Figure S7).  We define the \textit{anti--parallel} urea mole fraction as,
\begin{equation}
x_{\rm u}^{\rm anti-parallel} = \frac{N_{\rm u-r}^{\rm anti-parallel}}{N_{\rm u-r}^{\rm anti-parallel}+N_{\rm w-r}^{\rm anti-parallel}}\, ,
\end{equation}
which we use to compare with the fraction of H-bonds $x_{\rm u}^{\rm H-bond}$. To this end, we compute the ratio $\delta^{\rm anti-parallel}_{\rm H-bond}=x_{\rm u}^{\rm anti-parallel}/x_{\rm u}^{\rm H-bond}$. The data is shown in the lower panel of Figure~\ref{Figure_04}. 
Consistent with the energy--based arguments presented above, the plot contains two linear regimes with a crossover between $c_{\rm u}=$4 and 6~M. For $c_{\rm u} < 5$ M, the antiparallel dipole alignment is more important than H--bonds. However, when $c_{\rm u} > 5.0$ M, this picture gradually changes with H-bond formation becoming increasingly dominant. More interestingly, at $c_{\rm u}\approx$ 4 -- 5~M, there is a significant change in the behaviour of $\delta^{\rm antiparallel}_{\rm H-bond}$ with $c_{\rm u}$, separating regions with either dominant antiparallel arrangement or H-bond formation. This trend agrees with the re--entrant behaviour observed in Figure~\ref{Figure_01} that exhibits an apparent decrease in $\alpha-$helix formation at this concentration.\\

\section{Conclusions}

We have investigated the structure--thermodynamics relationship of poly--alanine in aqueous urea mixtures using molecular dynamics simulations of an all--atom model.
We emphasize the importance of a delicate competition between the H--bond and the dipole-dipole interactions that governs the macroscopic conformational behavior of the polypeptide sequence.
We provide a probable microscopic picture of the observations discussed in the experimental literature \cite{Scholtz_etal_PNAS92_185_1995,Hilser}. Our interpretation highlights that the effective models, such as the linear extrapolation method~\cite{Santoro_Bolen_Biochemistry27_8063_1988}, must be adjusted to explain the polypeptide unfolding as they severely underestimate the shift in solvation free energy \cite{Zhao:2020}. 
This work also challenges the common understanding based solely on the preferential interaction of urea with protein via vdW forces and shows that several delicate microscopic details, such as the local dipole--dipole interactions, are responsible for the 
polypeptide (or proteins in general) solvation in binary mixtures. In a broad sense, our results hint at a direct route to the operational understanding of the polypeptide interactions
in binary solutions. Thus, they may provide a new twist to our present understanding of protein solvation.
\begin{acknowledgement}
The authors thank Aysenur Iscen for a critical reading of the manuscript. L.A.B., K.K. and R.C.-H. gratefully acknowledge funding from SFB-TRR146 of the German Research Foundation (DFG). This project received funding from the European Research Council (ERC) under the European Union’s Seventh Framework Programme  (FP7/2007-2013)/ERC  Grant  Agreement No. 340906-MOLPROCOMP. D.M. thanks the Canada First Research Excellence Fund (CFREF), Quantum Materials and Future Technologies Program. Simulations have been performed on the THINC cluster at the Max Planck Institute for Polymer Research and on the COBRA cluster of the Max Planck Computing and Data Facility.
\end{acknowledgement}

\begin{suppinfo}
The Supplementary information contains four sections organized as follows. In Section S1, we present the model system and corresponding simulation details. The force field, with a particular focus on urea parameters and their validation, is presented in Section S2. Correlation functions taking into account the geometry of the problem, are defined and presented in Section S3. Finally, the dipole interaction model used to interpret our results is presented in Section S4. 	
\end{suppinfo}


\providecommand{\latin}[1]{#1}
\makeatletter
\providecommand{\doi}
  {\begingroup\let\do\@makeother\dospecials
  \catcode`\{=1 \catcode`\}=2 \doi@aux}
\providecommand{\doi@aux}[1]{\endgroup\texttt{#1}}
\makeatother
\providecommand*\mcitethebibliography{\thebibliography}
\csname @ifundefined\endcsname{endmcitethebibliography}
  {\let\endmcitethebibliography\endthebibliography}{}

\end{document}